\documentclass[useAMS,usenatbib]{mn2e}

\usepackage{graphicx}
\usepackage{amssymb}

\usepackage{natbib,times}
\usepackage{lscape}

\newcommand{\mc}{\multicolumn}

\begin{document}

\title[Mass Function of Rich Galaxy Clusters]
{Mass Function of Rich Galaxy Clusters and Its Constraint on $\sigma_8$}

\author[Z. L. Wen et al.]
{Z. L. Wen$^1$\thanks{E-mail: zhonglue@nao.cas.cn}, 
J. L. Han$^1$ 
and F. S. Liu$^2$
\\
$^1$National Astronomical Observatories, Chinese Academy of Sciences, 
20A Datun Road, Chaoyang District, Beijing 100012, China \\ 
$^2$College of Physics Science and Technology, Shenyang Normal University, 
Shenyang 110034, China \\
}

\date{Accepted 2009 ... Received 2009 ...}

\pagerange{\pageref{firstpage}--\pageref{lastpage}} \pubyear{2009}

\maketitle

\label{firstpage}


\begin{abstract}

The mass function of galaxy clusters is a powerful tool to constrain
cosmological parameters, e.g., the mass fluctuation on the scale of
8~$h^{-1}$~Mpc, $\sigma_8$, and the abundance of total matter,
$\Omega_m$. We first determine the scaling relations between cluster
mass and cluster richness, summed $r$-band luminosity and the global
galaxy number within a cluster radius. These relations are then used
to two complete volume-limited rich cluster samples which we obtained
from the Sloan Digital Sky Survey (SDSS). We estimate the masses of
these clusters and determine the cluster mass function. Fitting
the data with a theoretical expression, we get the cosmological
parameter constraints in the form of
$\sigma_8(\Omega_m/0.3)^{\alpha}=\beta$ and find out the parameters of
$\alpha=$0.40--0.50 and $\beta=$0.8--0.9, so that $\sigma_8=$0.8--0.9
if $\Omega_m=0.3$. Our $\sigma_8$ value is slightly higher than recent
estimates from the mass function of X-ray clusters and the 
Wilkinson Microwave Anisotropy Probe (WMAP) data, but consistent with
the weak lensing statistics.

\end{abstract}


\section{Introduction}

Precise determination of cosmological parameters is an important goal
in astrophysics. In the linear theory, the present root-mean-square
(rms) mass fluctuation on the scale of 8~$h^{-1}$~Mpc, $\sigma_8$, is
one of fundamental parameters \citep[see][]{svp+03} to describe the
power spectrum of mass fluctuations in the universe. It is one of key
parameters in the large scale structure simulations
\citep[e.g.,][]{jfp+98}.
The $\sigma_8$ can be determined by galaxy-galaxy correlations
\citep[e.g.,][]{tbs+04,cpp+05}, fluctuations in the cosmic microwave
background \citep{svp+03,sbd+07,kdn+09}, gravitational lensing
statistics \citep[e.g.,][]{hmv+06,kht+07,bhs+07}, cluster mass
function \citep[e.g.,][]{wef93,bf98,rb02}, Ly$\alpha$ forest
\citep{jnt+05,msc+05} and galaxy peculiar velocities \citep{fjf+03}.

The cluster mass function can be determined by the estimated masses
for a sample of clusters \citep[e.g.,][]{dah06}, or by the X-ray
luminosity and temperature function with a prior scaling relation
\citep{vl96,asf+03}. Fitting the cluster mass function with a
theoretical expression can provide constraint on
$\sigma_8$. Generally, $\sigma_8$ is coupled with $\Omega_m$, the
abundance of present total matter, in the form of
$\sigma_8(\Omega_m/0.3)^{\alpha}=\beta$. Previous studies have found
$\alpha$ in the range 0.3--0.6 and $\beta$ in the range 0.6--1.2 (see
Table~\ref{comp} in Section~\ref{discu}).
The determined $\sigma_8$ in recent years (2002--2009) has a mean
value of 0.73$\pm$0.05 assuming $\Omega_m=0.3$, which is in agreement
with the WMAP data \citep{kdn+09}, but lower than those by weak lensing
statistics \citep{hss+07}, galaxy-galaxy correlations
\citep{tbs+04,cpp+05} and Ly$\alpha$ forest \citep{jnt+05,msc+05}.

The amplitude of cluster mass function has large uncertainties, 
mainly caused by the uncertain normalization of the mass scaling
relation \citep[e.g., ][]{hen04}. Other uncertainties come from the
scatter of mass scaling relation and the incompleteness of the X-ray
flux-limited cluster samples \citep{rb02}. The cluster mass function 
may be underestimated if only X-ray clusters are used. \citet{evm+00} and
\citet{dpl+03} have noticed the existence of a class of
X-ray-underluminous massive clusters. \citet{pbb+07a} found that 40\%
of Abell clusters have a low level or no detection in
X-rays. 
A large complete volume-limited sample of clusters is crucial for the
purpose. Using the photometric redshifts of galaxies, we found 39,668
clusters in the redshift range $0.05<z<0.6$ \citep{whl09}. Clusters are
approximate volume-limited complete in the redshift range
$0.05<z<0.42$. The richnesses and the summed luminosities of clusters
are estimated from their luminous members, and they are tightly
related to cluster mass.
In Section 2, we carefully determine the scaling relation for cluster
mass. In Section 3, we get the cluster mass function for a local
sample of clusters and a sample at mediate redshifts, and then fit the
cluster mass function with a theoretical expression for constraints on
cosmological parameters, $\Omega_m$ and $\sigma_8$. Discussions and
conclusions are given in Section 4.

Throughout this paper, we assume a flat $\Lambda$CDM cosmology, taking
$H_0=$100 $h$ ${\rm km~s}^{-1}$ ${\rm Mpc}^{-1}$, with $h=0.72$,
$\Omega_m=1-\Omega_{\Lambda}$.

\begin{table*}
\begin{minipage}{160mm}
\caption[]{Cluster masses from literature and the mass tracer values for 53 clusters 
(richness $R\ge8$ and $0.03\le z\le0.3$, sorted with $z$) in the field of the SDSS DR6.}
\begin{center}
\begin{tabular}{lrrccccrlr}
\hline
\mc{1}{l}{Name}&\mc{1}{c}{R.A.}&\mc{1}{c}{Decl.}&\mc{1}{c}{$z$}&\mc{1}{c}{$R$}&
\mc{1}{c}{$L_r$} &\mc{1}{c}{$GGN/r_{\rm GGN}$}& \mc{1}{c}{M$_{\rm vir}$} & \mc{1}{c}{Method} & \mc{1}{c}{Ref.} \\
  &\mc{1}{c}{(deg)}& \mc{1}{c}{(deg)} & & &\mc{1}{c}{($10^{10}~h^{-2}~L_{\odot}$)} & \mc{1}{c}{(Mpc$^{-1}$)} &
\mc{1}{c}($10^{14}~h^{-1}~M_{\odot}$) & &\\
\hline
Abell 2199     &   247.15930 &   39.55121 &  0.030 & 17.76 & 28.49 & 18.10 &   $2.39^{+0.38}_{-0.38}$ &X-ray &  1 \\
               &             &            &        &       &       &       &   $3.42^{+0.26}_{-0.26}$ &X-ray &  2 \\
Abell 2052     &   229.18536 &    7.02162 &  0.035 &  9.50 & 21.97 & 11.62 &   $1.47^{+0.28}_{-0.28}$ &X-ray &  1 \\
               &             &            &        &       &       &       &   $1.58^{+0.05}_{-0.06}$ &X-ray &  2 \\
Abell 2063     &   230.77209 &    8.60922 &  0.035 & 15.30 & 22.96 & 19.93 &   $2.31^{+0.18}_{-0.16}$ &X-ray &  2 \\
               &             &            &        &       &       &       &   $2.78^{+0.42}_{-0.42}$ &X-ray &  1 \\
Abell 2147     &   240.57094 &   15.97465 &  0.035 & 12.22 & 22.50 & 14.67 &   $2.46^{+0.83}_{-0.52}$ &X-ray &  2 \\
Abell 2151w    &   241.14914 &   17.72156 &  0.037 & 11.08 & 22.16 & 13.29 &   $1.23^{+0.09}_{-0.09}$ &X-ray &  2 \\
MKW9           &   233.13339 &    4.68100 &  0.040 &  8.88 & 15.77 & 11.44 &   $1.01^{+0.25}_{-0.53}$ &X-ray &  3 \\
               &             &            &        &       &       &       &   $1.06^{+0.27}_{-0.27}$ &X-ray &  4 \\
Abell 1983     &   223.23048 &   16.70286 &  0.044 &  9.27 & 11.76 & 11.45 &   $1.39^{+0.53}_{-0.53}$ &X-ray &  3 \\
               &             &            &        &       &       &       &   $1.41^{+0.55}_{-0.55}$ &X-ray &  4 \\
Abell 160      &    18.24822 &   15.49129 &  0.045 & 16.61 & 34.96 & 16.54 &   $0.96^{+0.13}_{-0.14}$ &X-ray &  5 \\
Abell 85       &    10.46029 & $-$9.30312 &  0.056 & 23.61 & 54.65 & 21.85 &   $4.23^{+1.15}_{-1.15}$ &X-ray &  1 \\
               &             &            &        &       &       &       &   $5.55^{+0.58}_{-0.53}$ &X-ray &  2 \\
Abell 1991     &   223.63122 &   18.64232 &  0.059 & 19.22 & 39.13 & 17.48 &   $1.35^{+0.19}_{-0.53}$ &X-ray &  3 \\
               &             &            &        &       &       &       &   $1.44^{+0.17}_{-0.17}$ &X-ray &  4 \\
               &             &            &        &       &       &       &   $1.52^{+0.21}_{-0.21}$ &X-ray &  6 \\
               &             &            &        &       &       &       &   $1.66^{+0.32}_{-0.24}$ &X-ray &  5 \\
Abell 1795     &   207.21877 &   26.59293 &  0.063 & 19.56 & 39.70 & 19.56 &   $5.42^{+0.65}_{-0.65}$ &X-ray &  1 \\
               &             &            &        &       &       &       &   $7.58^{+1.92}_{-1.70}$ &X-ray &  7 \\
               &             &            &        &       &       &       &   $7.86^{+0.70}_{-0.70}$ &X-ray &  6 \\
               &             &            &        &       &       &       &   $7.94^{+1.64}_{-1.51}$ &X-ray &  2 \\
Abell 2092     &   233.31403 &   31.14515 &  0.067 &  9.27 & 15.93 & 11.45 &   $1.09^{+0.23}_{-0.20}$ &X-ray &  5 \\
Abell 2065     &   230.60008 &   27.71436 &  0.072 & 40.86 & 76.31 & 40.86 &  $12.11^{+15.48}_{-4.88}$&X-ray &  2 \\
ZwCl 1215      &   184.42134 &    3.65584 &  0.075 & 20.69 & 43.81 & 18.71 &   $7.54^{+2.55}_{-1.91}$ &X-ray &  2 \\
Abell 1800     &   207.34822 &   28.10732 &  0.075 & 19.77 & 44.73 & 18.83 &   $4.14^{+4.31}_{-1.65}$ &X-ray &  2 \\
Abell 1775     &   205.45477 &   26.37347 &  0.076 & 14.52 & 32.62 & 17.76 &   $3.07^{+0.43}_{-0.29}$ &X-ray &  2 \\
Abell 2029     &   227.73376 &    5.74478 &  0.077 & 24.87 & 60.63 & 27.47 &   $7.06^{+0.69}_{-0.54}$ &X-ray &  7 \\
               &             &            &        &       &       &       &   $9.30^{+1.93}_{-1.93}$ &X-ray &  1 \\
               &             &            &        &       &       &       &   $9.76^{+1.77}_{-1.65}$ &X-ray &  2 \\
               &             &            &        &       &       &       &  $10.55^{+1.01}_{-1.01}$ &X-ray &  6 \\
Abell 2255     &   258.11996 &   64.06072 &  0.080 & 40.23 & 83.40 & 33.51 &   $9.70^{+1.05}_{-0.87}$ &X-ray &  2 \\
Abell 1650     &   194.67288 & $-$1.76146 &  0.084 & 17.08 & 42.92 & 20.68 &   $8.13^{+4.21}_{-2.53}$ &X-ray &  2 \\
Abell 1692     &   198.05661 & $-$0.97448 &  0.085 & 14.40 & 26.42 & 14.73 &   $1.19^{+0.38}_{-0.24}$ &X-ray &  5 \\
Abell 1750     &   202.71080 & $-$1.86197 &  0.088 & 24.35 & 58.05 & 27.58 &   $4.78^{+2.64}_{-2.64}$ &WL    &  8 \\
Abell 2142     &   239.58334 &   27.23341 &  0.090 & 39.33 & 80.90 & 39.88 &  $11.00^{+2.85}_{-1.93}$ &X-ray &  2 \\
               &             &            &        &       &       &       &  $11.95^{+5.24}_{-5.24}$ &WL    &  8 \\
Abell 2244     &   255.67705 &   34.05999 &  0.097 & 31.22 & 62.35 & 28.67 &   $7.23^{+9.45}_{-3.25}$ &X-ray &  2 \\
Abell 2034     &   227.54883 &   33.48646 &  0.113 & 32.00 & 68.97 & 31.08 &   $7.17^{+4.30}_{-4.30}$ &WL    &  8 \\
Abell 1068     &   160.18541 &   39.95313 &  0.138 & 16.38 & 38.87 & 16.38 &   $4.87^{+0.42}_{-0.53}$ &X-ray &  3 \\
               &             &            &        &       &       &       &   $5.12^{+0.45}_{-0.45}$ &X-ray &  4 \\
Abell 1413     &   178.82501 &   23.40491 &  0.143 & 37.17 & 93.39 & 41.82 &   $5.70^{+0.57}_{-0.53}$ &X-ray &  3 \\
               &             &            &        &       &       &       &   $5.87^{+0.59}_{-0.59}$ &X-ray &  4 \\
               &             &            &        &       &       &       &   $8.72^{+0.80}_{-0.70}$ &X-ray &  2 \\
               &             &            &        &       &       &       &   $9.76^{+2.54}_{-1.89}$ &X-ray &  7 \\
               &             &            &        &       &       &       &   $9.95^{+1.04}_{-1.04}$ &X-ray &  6 \\
RXJ1720.1$+$2637&  260.04184 &   26.62557 &  0.164 & 26.87 & 69.11 & 27.30 &   $5.88^{+4.98}_{-4.60}$ &WL    &  9 \\
               &             &            &        &       &       &       &   $5.84^{+4.30}_{-4.30}$ &WL    & 10 \\
Abell 1914     & 216.48611   & 37.81645  & 0.171  & 34.16 &  84.15 &  38.07  & $ 4.66^{+3.78}_{ -3.49}$ &WL   &  9 \\
               &             &           &        &	 &        &         & $ 6.14^{+3.19}_{ -3.19}$ &WL   &  8 \\
               &             &           &        &	 &        &         & $ 9.08^{+5.66}_{ -5.66}$ &WL   & 10 \\
               &             &           &        &	 &        &         & $18.40^{+2.20}_{ -1.86}$ &X-ray&  2 \\
MS 0906.5$+$1110& 137.30312  & 10.97475  & 0.176  & 36.00 & 119.08 &  38.61  & $ 8.30^{+2.30}_{ -2.30}$ &WL   & 11 \\
\end{tabular}
\end{center}
\label{mass}
\end{minipage}
\end{table*}

\addtocounter{table}{-1}
\begin{table*}
\begin{minipage}{160mm}
\caption[]{-- {\it continued} }
\begin{center}
\begin{tabular}{lrrccccrlr}
\hline
\mc{1}{l}{Name}&\mc{1}{c}{R.A.}&\mc{1}{c}{Decl.}&\mc{1}{c}{$z$}&\mc{1}{c}{$R$}&
\mc{1}{c}{$L_r$} &\mc{1}{c}{$GGN/r_{\rm GGN}$}& \mc{1}{c}{M$_{\rm vir}$} & \mc{1}{c}{Method} & \mc{1}{c}{Ref.} \\
  &\mc{1}{c}{(deg)}& \mc{1}{c}{(deg)} & & &\mc{1}{c}{($10^{10}~h^{-2}~L_{\odot}$)} & \mc{1}{c}{(Mpc$^{-1}$)} &
\mc{1}{c}($10^{14}~h^{-1}~M_{\odot}$) & &\\
\hline
Abell 1689     & 197.87291  &$-$1.34108 & 0.184  & 62.74 & 136.85 &  50.32  & $10.50^{+1.91}_{ -1.91}$ &WL   & 12 \\
               &            &           &        &	 &        &         & $11.69^{+1.80}_{ -1.80}$ &SL+WL& 13 \\
               &            &           &        &	 &        &         & $13.24^{+3.99}_{ -3.99}$ &X-ray& 14 \\
               &            &           &        &	 &        &         & $13.43^{+0.93}_{ -0.96}$ &X-ray&  2 \\
               &            &           &        &	 &        &         & $13.51^{+1.40}_{ -1.40}$ &WL   & 15 \\
               &            &           &        &	 &        &         & $14.70^{+1.40}_{ -1.40}$ &WL   & 16 \\
               &            &           &        &	 &        &         & $17.55^{+3.04}_{ -3.04}$ &WL   & 17 \\
               &            &           &        &	 &        &         & $20.53^{+1.74}_{ -1.74}$ &WL   & 18 \\
               &            &           &        &	 &        &         & $25.50^{+5.30}_{ -4.50}$ &WL   & 11 \\
Abell 963      & 154.26515  & 39.04705  & 0.206  & 43.00 &  93.83 &  38.52  & $ 3.45^{+0.80}_{ -0.80}$ &WL   & 17 \\
               &            &           &        &	 &        &         & $ 6.50^{+2.00}_{ -1.90}$ &WL   & 11 \\
               &            &           &        &	 &        &         & $ 6.53^{+2.00}_{ -2.00}$ &X-ray& 14 \\
               &            &           &        &	 &        &         & $ 7.00^{+2.45}_{ -1.60}$ &X-ray&  7 \\
               &            &           &        &	 &        &         & $ 8.01^{+8.14}_{ -6.36}$ &WL   &  9 \\
               &            &           &        &	 &        &         & $ 8.38^{+7.46}_{ -7.46}$ &WL   & 10 \\
Abell 1423     & 179.32219  & 33.61092  & 0.214  & 20.78 &  64.82 &  21.16  & $14.98^{+7.67}_{ -7.67}$ &WL   & 10 \\
RX J1504.1$-$0248&226.03130 &$-$2.80460 & 0.215  & 20.08 &  73.32 &  26.82  & $15.10^{+9.40}_{ -4.30}$ &X-ray&  7 \\
Abell 773      & 139.47261  & 51.72704  & 0.217  & 66.83 & 138.50 &  55.08  & $10.63^{+3.27}_{ -3.27}$ &X-ray& 14 \\
               &            &           &        &	 &        &         & $15.33^{+6.95}_{ -6.95}$ &WL   & 10 \\
               &            &           &        &	 &        &         & $24.70^{+9.44}_{-11.91}$ &WL   &  9 \\
Abell 1682     & 196.70833  & 46.55927  & 0.226  & 40.76 & 112.14 &  39.49  & $ 3.96^{+3.52}_{ -2.40}$ &WL   &  9 \\
               &            &           &        &	 &        &         & $ 5.49^{+3.72}_{ -3.72}$ &WL   & 10 \\
Abell 1763     & 203.83372  & 41.00115  & 0.228  & 34.10 & 109.20 &  32.77  & $ 6.23^{+1.92}_{ -1.92}$ &X-ray& 14 \\
               &            &           &        &       &        &         & $ 8.91^{+4.60}_{ -5.70}$ &WL   &  9 \\
               &            &           &        &	 &        &         & $10.54^{+5.16}_{ -5.16}$ &WL   & 10 \\
               &            &           &        &	 &        &         & $12.28^{+2.37}_{ -2.37}$ &WL   & 17 \\
               &            &           &        &       &        & 	    & $13.50^{+3.70}_{ -3.30}$ &WL   & 11 \\
Abell 2219     & 250.08253  & 46.71148  & 0.228  & 45.62 &  94.82 &  45.62  & $ 8.19^{+5.69}_{ -5.69}$ &WL   & 10 \\
               &            &           &        &	 &        &         & $ 8.57^{+4.46}_{ -5.45}$ &WL   &  9 \\
               &            &           &        &	 &        &         & $11.30^{+3.20}_{ -2.70}$ &WL   & 11 \\
               &            &           &        &	 &        &         & $18.66^{+3.94}_{ -3.94}$ &WL   & 17 \\
Abell 2111     & 234.91872  & 34.42426  & 0.229  & 47.58 & 112.27 &  43.38  & $ 6.12^{+3.64}_{ -3.64}$ &WL   & 10 \\
               &            &           &        &       &        &         & $ 6.92^{+3.39}_{ -3.95}$ &WL   &  9 \\
Abell 267      &  28.17483  &  1.00711  & 0.230  & 31.53 &  86.85 &  32.62  & $ 2.36^{+1.29}_{ -1.29}$ &WL   & 17 \\
               &            &           &        &       &        &         & $ 5.36^{+1.70}_{ -1.70}$ &X-ray& 14 \\
               &            &           &        &	 &        &         & $ 7.50^{+2.30}_{ -2.20}$ &WL   & 11 \\
               &            &           &        &	 &        &         & $14.98^{+5.63}_{ -5.63}$ &WL   & 10 \\
               &            &           &        &	 &        &         & $16.34^{+5.57}_{ -6.75}$ &WL   &  9 \\
Zw 1231.4$+$1007 & 188.57277  &  9.76623  & 0.231  & 32.00 &  98.28 &  36.99  & $ 3.15^{+2.65}_{ -2.65}$ &WL   & 10 \\
MS 1231.3$+$1542 & 188.48055  & 15.43305  & 0.234  & 13.82 &  35.53 &  17.40  & $ 1.50^{+0.90}_{ -0.90}$ &WL   & 11 \\
RX J2129.6$+$0005& 322.41649  &  0.08920  & 0.234  & 24.00 &  80.65 &  20.45  & $ 5.96^{+8.54}_{ -2.70}$ &X-ray&  7 \\
               &            &           &        &	 &        &         & $ 7.51^{+5.24}_{ -5.24}$ &WL   & 10 \\
               &            &           &        &       &        &         & $15.49^{+8.97}_{ -9.24}$ &WL   &  9 \\
Abell 1835     & 210.25863  &  2.87846  & 0.252  & 51.57 & 129.59 &  47.96  & $ 8.19^{+3.07}_{ -3.07}$ &WL   & 12 \\
               &            &           &        &	 &        &         & $ 8.41^{+2.57}_{ -2.57}$ &X-ray& 14 \\
               &            &           &        &	 &        &         & $10.61^{+5.69}_{ -5.69}$ &WL   & 10 \\
               &            &           &        &	 &        &         & $15.62^{+8.56}_{ -6.35}$ &WL   &  9 \\
               &            &           &        &	 &        &         & $17.00^{+3.10}_{ -3.40}$ &X-ray&  7 \\
               &            &           &        &	 &        &         & $24.21^{+3.76}_{ -3.76}$ &WL   & 17 \\
MS 1455.0$+$2232 & 224.31295  & 22.34288  & 0.258  & 26.00 &  72.03 &  26.00  & $ 5.75^{+4.11}_{ -3.26}$ &WL   &  9 \\
               &            &           &        &	 &        &         & $ 7.30^{+1.90}_{ -1.80}$ &WL   & 11 \\
               &            &           &        &	 &        &         & $12.88^{+6.07}_{ -6.07}$ &WL   & 10 \\
Abell 2631     & 354.41554  &  0.27138  & 0.277  & 46.22 & 113.32 &  44.94  & $ 6.12^{+4.25}_{ -4.25}$ &WL   & 10 \\
Abell 1758N    & 203.16007  & 50.55992  & 0.279  & 42.24 & 107.00 &  41.36  & $ 5.26^{+5.70}_{ -5.70}$ &WL   &  8 \\
               &            &           &        &       &        &         & $26.91^{+9.57}_{ -9.57}$ &WL   & 10 \\
               &            &           &        &       &        &         & $39.09^{+12.77}_{-13.16}$&WL   &  9 \\
\end{tabular}
\end{center}
\end{minipage}
\end{table*}

\addtocounter{table}{-1}
\begin{table*}
\begin{minipage}{160mm}
\caption[]{-- {\it continued} }
\begin{center}
\begin{tabular}{lrrccccrlr}
\hline
\mc{1}{l}{Name}&\mc{1}{c}{R.A.}&\mc{1}{c}{Decl.}&\mc{1}{c}{$z$}&\mc{1}{c}{$R$}&
\mc{1}{c}{$L_r$} &\mc{1}{c}{$GGN/r_{\rm GGN}$}& \mc{1}{c}{M$_{\rm vir}$} & \mc{1}{c}{Method} & \mc{1}{c}{Ref.} \\
  &\mc{1}{c}{(deg)}& \mc{1}{c}{(deg)} & & &\mc{1}{c}{($10^{10}~h^{-2}~L_{\odot}$)} & \mc{1}{c}{(Mpc$^{-1}$)} &
\mc{1}{c}($10^{14}~h^{-1}~M_{\odot}$) & &\\
\hline
Abell 697      & 130.73982  & 36.36646  & 0.282  & 27.72 &  77.66 &  37.83  & $22.92^{+9.78}_{ -9.46}$ &WL   &  9 \\
               &            &           &        &       &        &         & $26.10^{+9.54}_{ -9.54}$ &WL   & 10 \\
Abell 959      & 154.39984  & 59.56710  & 0.285  & 52.77 & 139.82 &  48.82  & $17.38^{+6.81}_{ -5.85}$ &WL   &  9 \\
Abell 611      & 120.23674  & 36.05655  & 0.288  & 27.86 &  91.23 &  30.92  & $ 6.18^{+3.82}_{ -1.81}$ &X-ray&  7 \\
               &            &           &        &       &        &         & $ 6.54^{+4.43}_{ -4.43}$ &WL   & 10 \\
               &            &           &        &       &        &         & $ 6.90^{+5.65}_{ -5.11}$ &WL   &  9 \\
Abell 781      & 140.20117  & 30.47176  & 0.288  & 27.17 &  77.19 &  33.22  & $12.67^{+5.86}_{ -5.86}$ &WL   & 10 \\
Zw3146         & 155.91515  &  4.18629  & 0.291  & 31.76 &  70.81 &  30.33  & $11.31^{+5.44}_{ -5.44}$ &WL   & 10 \\
               &            &           &        &       &        &         & $13.99^{+7.99}_{ -6.17}$ &WL   &  9 \\
Zw1459.4+4240  & 225.34604  & 42.34448  & 0.292  & 27.67 &  81.26 &  29.89  & $10.59^{+6.00}_{ -6.00}$ &WL   & 10 \\
Abell 1576     & 189.24684  & 63.18658  & 0.300  & 42.00 & 156.45 &  38.40  & $16.01^{+6.60}_{ -4.86}$ &WL   &  9 \\
               &            &           &        &       &        &         & $18.45^{+6.03}_{ -6.03}$ &WL   & 10 \\
\hline
\end{tabular}
\end{center}
{
Note for Method: X-ray stands for the mass determined by X-ray measurements; WL stands for weak lensing, 
WL+SL stands for weak lensing combined with strong lensing; References for mass estimates: 
[1] \citet{xjw01}, [2] \citet{rb02}, [3] \citet{pap05}, [4] \citet{app05}, 
[5] \citet{svd+09}, [6] \citet{vkf+06}, [7] \citet{sa07}, [8] \citet{ou08}, [9] \citet{pd07}, 
[10] \citet{dah06}, [11] \citet{hoe07}, [12] \citet{ckc09}, [13] \citet{lrj+07}, [14] \citet{zfb+07}, [15] \citet{btu+05}, 
[16] \citet{ub08}, [17] \citet{bsk+07}, [18] \citet{hsp06}.
} 
\end{minipage}
\end{table*}  

\begin{figure*}
\centering
\includegraphics[width = 17.cm]{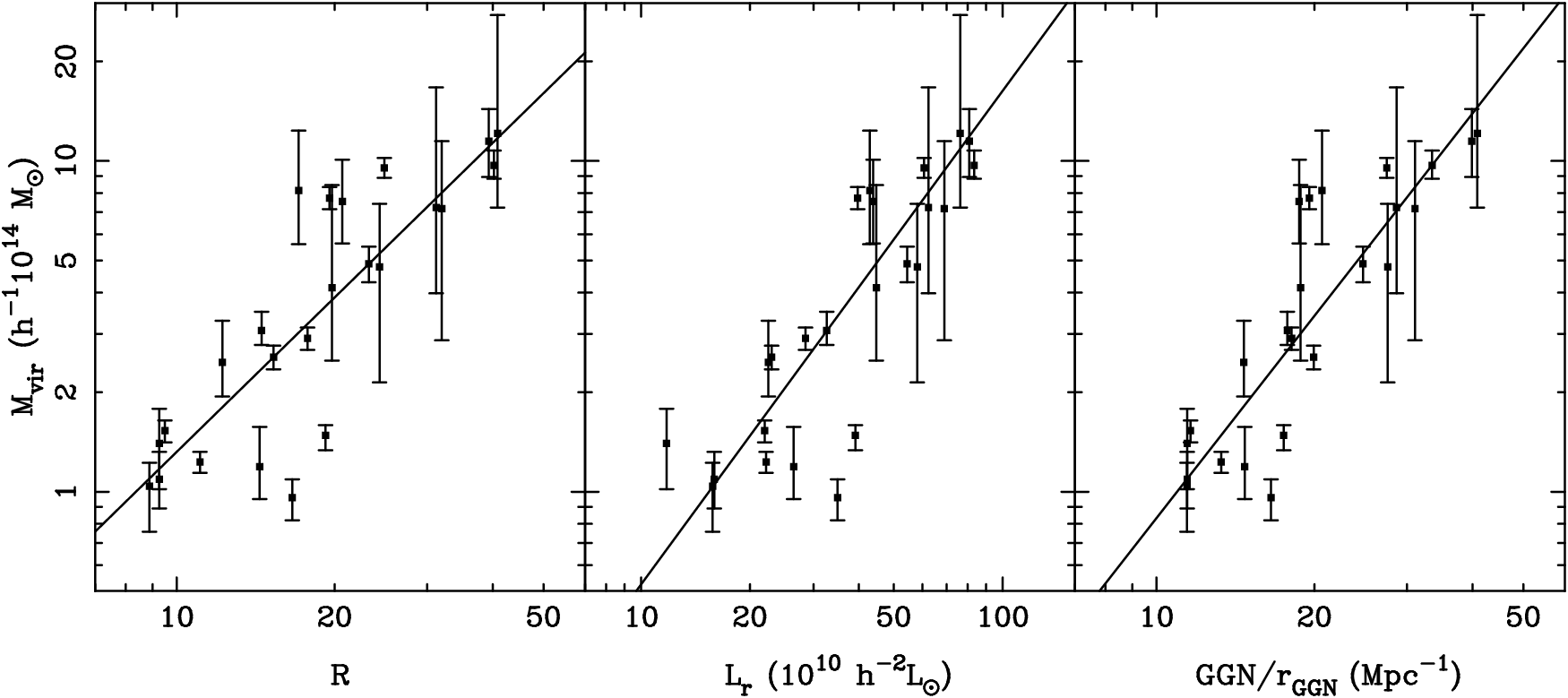}
\caption{Correlations between cluster mass $M_{\rm vir}$ and richness $R$, summed luminosity $L_r$ and
$GGN/r_{GGN}$ for 24 nearby clusters ($R\ge8$ and $z\lesssim0.1$). The solid 
line is the best fit as given in Equation~(\ref{mr})--(\ref{mggn}).}
\label{richmass}
\end{figure*}

\begin{figure*}
\centering
\includegraphics[width = 17.cm]{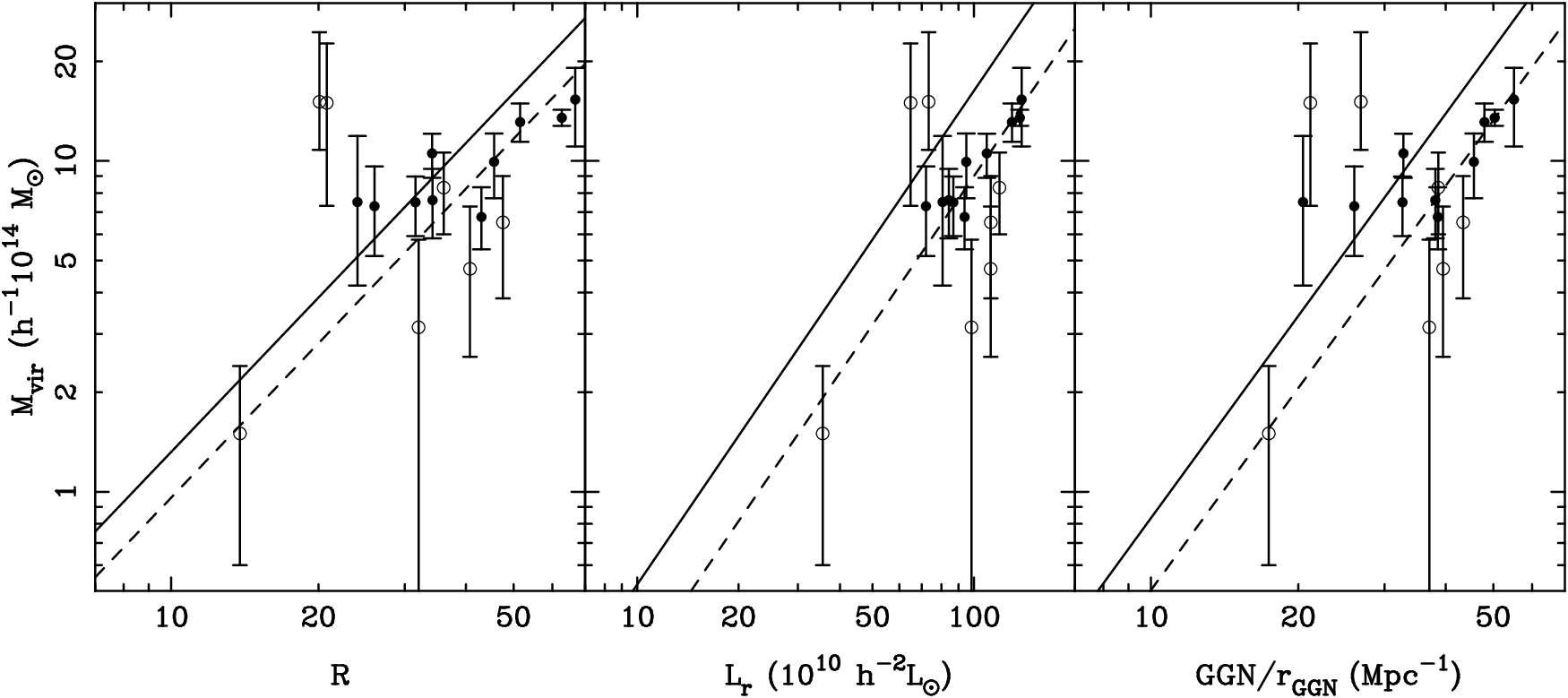}
\caption{Correlations between cluster mass $M_{\rm vir}$ and richness $R$, summed
luminosity $L_r$ and $GGN/r_{GGN}$ for 17 clusters in the redshift range
$0.17<z<0.26$. The black dots are the clusters with more than three
estimates of their masses. The solid lines are the same shown 
in Figure~\ref{richmass}. The dashed line is the new
scaling relation with the same slope but
different offsets determined from the data.}
\label{richmass2}
\end{figure*}

\section{Mass scaling relations for clusters}

We identified 39,668 clusters from the SDSS DR6 by 
discrimination of luminous member galaxies with following steps \citep{whl09}. 
First, we assume that each galaxy at a given photometric
redshift $z$ is the central galaxy of a cluster candidate, and
we count the number of luminous ``member galaxies'' of $M_r\le-21$
within a radius of 0.5 Mpc and a photometric redshift gap of 
$z\pm0.04(1+z)$. We set $\Delta z=0.04(1+z)$ for the gap to allow variable
uncertainties of photometric redshifts at different redshifts.
Second, we define the center of a cluster candidate to be the
position of the galaxy with a maximum number count.  The cluster
redshift is estimated to be the median value of the photometric
redshifts of the recognized ``members''. Third, for each cluster
candidate at $z$, all galaxies within 1 Mpc from the cluster center
and $z\pm0.04(1+z)$ are assumed
to be the member galaxies. Their absolute magnitudes are
re-calculated with the cluster redshift. Finally, a cluster at $z$ is
identified when the number of member galaxies of $M_r\le-21$ 
reaches 8 within a projected radius of 0.5 Mpc and $z\pm\Delta z$. 
Monte-Carlo simulations show that the detection rate is more than
90\% for massive clusters (richness $R\ge16.7$) if the redshift 
uncertainty of cluster galaxies is about $0.03(1+z)$.

We defined the cluster richness, $R$, to be the total number of
galaxies ($M_r\le-21$) within a radius of 1 Mpc and $z\pm
0.04(1+z)$ after subtracting the local
background, i.e., the average number of luminous galaxies. 
The summed $r$-band luminosity of each cluster, $L_r$, is
calculated as the total luminosity of member galaxies within the
region also after subtracting the background. From the radial distribution
of member galaxies, we got the cluster radius, $r_{GGN}$, where the
density of galaxies is as low as background. Here, we defined the 
Gross Galaxy Number ($GGN$) of a cluster as the total number of luminous galaxies ($M_r\le-21$) within 
the radius $r_{GGN}$ and the redshift gap of $z\pm 0.04(1+z)$ after subtracting the local background. 
It has been known for a long time
that the cluster richness and summed luminosity are related to cluster
mass \citep{gmm+02,pbb+07b}, hence they can be the tracers of cluster
mass. The $GGN/r_{GGN}$ is related to the amplitude of
cluster-galaxy cross-correlation since the correlation is described by
$\xi (r)\propto r^{-2}$ \citep[e.g.,][]{le88}.  We find that
$GGN/r_{GGN}$ can also be the tracer of cluster richness.

Cluster mass can be determined by the velocity dispersion of member
galaxies \citep{zwi33}. However, velocity measurements can be
corrupted by projection effects that might be difficult to diminish in
practice. The error on the individual measurements can introduce a
significant bias \citep{vbk+07}.
Under the assumption of hydrostatic equilibrium, the X-ray method can
determine mass distribution of a cluster to a large radius. The
assumption is invalid for clusters with substructures, inducing an
underestimation of mass \citep{sch96}. Weak gravitational lensing 
recently becomes a sophisticated method to estimate cluster 
mass without assumptions on dynamical state of a cluster. 
The uncertainty of mass mainly comes from the difficulty in measuring 
the image distortions of the faint background sources.
We collect the cluster masses estimated by X-ray and weak lensing
methods from literature (see Table~\ref{mass}). Usually, cluster
masses are denoted as $M_{\Delta}$ which is the mass within a radius
$r_{\Delta}$ interior to which the mean density is $\Delta$ times the
critical density of the universe. For cosmology with $\Omega_m=0.3$,
the virial mass is calculated within the radius $r_{\Delta}$, here
$\Delta=101$, so that $M_{\rm vir}=M_{101}$ \citep{ks96}. Previous
studies usually provided the mass within $r_{200}$ or $r_{500}$
\citep[e.g.,][]{rb02,pd07}. Here, we convert the mass of $M_{\rm 200}$
and $M_{\rm 500}$ to the virial mass $M_{\rm vir}$ according to
\citet{sks+03}. We will discuss later the influence on our result
from a possible bias conversion.
For each cluster with mass estimated, we calculate the cluster
richness, the summed $r$-band luminosity and $GGN/r_{GGN}$ following
the method of \citet{whl09}. Only clusters of richness $R\ge8$ are
listed in Table~\ref{mass} since the uncertainties of $R$ and the
summed luminosities become larger for clusters with a smaller $R$.

We notice that clusters with estimated masses preferentially have low
($z\lesssim0.1$) and mediate ($\sim0.2< z< 0.25$) redshifts (see
Table~\ref{mass}). To minimize the uncertainty, we determine the
scaling relations between the masses and observational tracers for
clusters in the two small redshift ranges independently. This is
because the discrimination of member galaxies (e.g., completeness or
contamination rate) may be different for clusters at different
redshifts, and the systematic bias can be ignored in such a small
range.
In the low redshift range ($z\lesssim0.1$), the masses of many
clusters are available and distributed in a large mass range, which is
good for determination of the scaling relations. We get 15 clusters of
$0.05<z<0.1$. We also include 8 clusters of $0.03<z<0.05$ and one cluster
of $z=0.113$ to derive the scaling relations at the low redshift
range. Several clusters have multiple estimates for mass from 
literature, we adopt the median value or the average of 
two middle ones for even measurements.

The mass--richness relation, i.e., the so called halo occupation
distribution in some literature \citep[e.g.,][]{pbb+07b}, is described
by a power law, $R\propto M^{\mu}$.  The correlation of cluster mass
with the optical luminosity, i.e., the mass-to-light ratio $M/L$, is
also described by a power law, $M/L\propto L^{\nu}$, i.e., $M\propto L^{1+\nu}$. In
Figure~\ref{richmass}, we show the correlations between cluster mass
and cluster richness, summed luminosity and $GGN/r_{GGN}$ for 24 nearby
clusters. The uncertainties of richness $R$, summed luminosity $L_r$ 
and $GGN/r_{GGN}$ are about 10\%--20\% \citep{whl09}. We fit the 
correlations with power-law relations,

\begin{equation}
\log M_{\rm vir}=(-1.43\pm0.07)+(1.55\pm0.06)\log R,
\label{mr}
\end{equation}
\begin{equation}
\log M_{\rm vir}=(-1.77\pm0.08)+(1.49\pm0.05)\log L_r,
\label{ml}
\end{equation}
and
\begin{equation}
\log M_{\rm vir}=(-2.11\pm0.10)+(2.03\pm0.08)\log (GGN/r_{GGN}).
\label{mggn}
\end{equation}
Here, $M_{\rm vir}$ has a unit of $10^{14}~h^{-1}~M_{\odot}$, $L_r$
has a unit of $10^{10}~h^{-2}~L_{\odot}$. The uncertainty of
the estimated cluster mass, $\sigma_{\log M}$, is mainly determined 
by the uncertainties of the intercept and the slope in the logarithm 
for three scaling relations in Equation~(\ref{mr})--(\ref{mggn}). \citet{ye03} defined
$B_{\rm gc}$ to be the amplitude of galaxy-cluster cross-correlation
function and found $M_{\rm vir}\propto B_{\rm gc}^{1.64\pm0.28}$. The
slope is in agreement with that of our $M_{\rm vir}$ to $GGN/r_{GGN}$
relation.  These scaling relations, Equation~(\ref{mr})--(\ref{mggn}),
will be used to estimate masses of a complete volume-limited sample of
clusters in the local universe for cluster mass function.

We can also use a much larger cluster sample at mediate redshift
($\sim0.2< z< 0.25$) for cluster mass function. Some massive clusters
in this redshift range have their masses estimated (see
Table~\ref{mass}).
We obtain masses of 17 clusters in the redshift range of $0.17< z<
0.26$, of which 10 clusters have more than three estimates.  In
Figure~\ref{richmass2}, we show the correlations between cluster mass
and cluster richness, summed luminosity and $GGN/r_{GGN}$ for the 17
clusters. Most of them are similarly massive of $10^{15}~h^{-1}~M_{\odot}$ 
and few have smaller masses, 
so that it is difficult to determine a new scaling relations. Here, we
calibrate the mass scaling relations by assuming the same slopes of
Equation~(\ref{mr})--(\ref{mggn}) and finding the offsets. We then get 
the scaling relations,

\begin{equation}
\log M_{\rm vir}=(-1.57\pm0.12)+1.55\log R,
\label{mr2}
\end{equation}
\begin{equation}
\log M_{\rm vir}=(-2.03\pm0.06)+1.49\log L_r,
\label{ml2}
\end{equation}
and
\begin{equation}
\log M_{\rm vir}=(-2.33\pm0.11)+2.03\log (GGN/r_{GGN}).
\label{mggn2}
\end{equation}
The uncertainties in Equation~(\ref{mr2})--(\ref{mggn2}) reflect the
scatters of masses to the mean relations (dashed line).  We notice
that the scatter is the smallest for the $M_{\rm vir}$--$L_r$ relation
for the high redshift data, because clusters with more than three
estimates (black dots) are very consistent with the fitting relation
(dashed line). Therefore, the cluster masses estimated by the $M_{\rm
  vir}$--$L_r$ relation may be more accurate than other tracers.
The offsets between the relations for samples at two redshift ranges
may come from the problem of the SDSS galaxy data. The sky background
level is overestimated for nearby bright galaxies ($12.5<r<15.5$), so
that galaxies have systematically fainter magnitudes by 0.15--0.2 mag
than their true magnitude \citep{dr6}. This can result in
systematically lower cluster richness and summed luminosity for
clusters of $0.05<z<0.1$ than clusters of $0.2<z<0.25$. The two
scaling relations are used to samples of clusters at two redshift
ranges independently. Hence, the systematic bias does not affect
the final $\sigma_8$ values from each sample. 

\section{Cluster mass functions}
\label{clustermf}

Assuming a Gaussian distribution of mass fluctuation, \citet{ps74}
used a linear theory to derive the first theoretical expression of
cluster mass function, which is in agreement with mass functions
derived from observations and numerical simulations within a large
mass range \citep[e.g.,][]{wef93,rb02}. Recent simulations show
slightly more massive clusters than the Press \& Schechter mass
function gives \citep{st99,jfw+01,wah+06}. In this work, we take the
form of the cluster mass function as Equation (B4) of
  \citet{jfw+01}. The mean differential comoving number density of
dark matter halos is

\begin{equation}
\frac{dn}{dM}=0.316\frac{\rho_0}{M^2}\frac{d \ln \sigma^{-1}}{d \ln M}\exp(-[\ln \sigma^{-1}+0.67]^{3.82}).
\label{jenk}
\end{equation}
Here, $\rho_0=2.78\times10^{11}\Omega_mh^2~M_{\odot}$~Mpc$^{-3}$ is
the comoving density of the universe. $M$ is the halo mass within a
radius with a mean overdensity of 324 times of the mean density of the
universe (roughly the virial mass, $M_{101}$, if $\Omega_m=0.3$).
$\sigma^2(M,z)$ is the variance of the linearly evolved density 
field smoothed by a spherical top-hat filter that enclose mass $M$. 
Here, $\sigma(M,z)=\sigma_8\times f$, where $\sigma_8$ is the 
present linear rms mass fluctuation on the scale of 8 $h^{-1}$ Mpc 
and $f$ is a function of $M$, $z$, $\Omega_m$ as well as the Hubble 
constant $h$, the abundance of baryons $\Omega_b$ and the present 
cosmic microwave background temperature $T_{\rm CMB}$. 
$d \ln \sigma^{-1}/d \ln M$ can be derived from the expression 
of $\sigma(M,z)$ \citep[see details of $\sigma(M,z)$ in][]{rb02}. 
The values of $\Omega_m$ and $\sigma_8$ are the main parameters 
to define the mass function. The other parameters does not 
strongly affect the results in our analysis, thus can be fixed. 
The $\sigma_8$ strongly depends on cluster mass function at the high
mass end. Since the mass function is steep at high mass end, 
the data scatter for mass scaling relations induces more low mass 
to higher mass. Thus, the uncertainty of the mass scaling relation, 
$\sigma_{\log M}$, is included in the fitting. We re-write the mass 
function with the uncertainty on mass estimate to be the Jenkins function 
convolved by a Gaussian function,
\begin{equation}
\frac{d\tilde{n}(M)}{d\log M}=\int \frac{dn(M')}{d\log M'}g(\log M-\log M',\sigma_{\log M})d\log M',
\label{jenk2}
\end{equation}
where $g(x,\sigma)=e^{-x^2/2\sigma^2}/(\sqrt{2\pi}\sigma)$.

\begin{figure*}
\centering
\includegraphics[width = 17.cm]{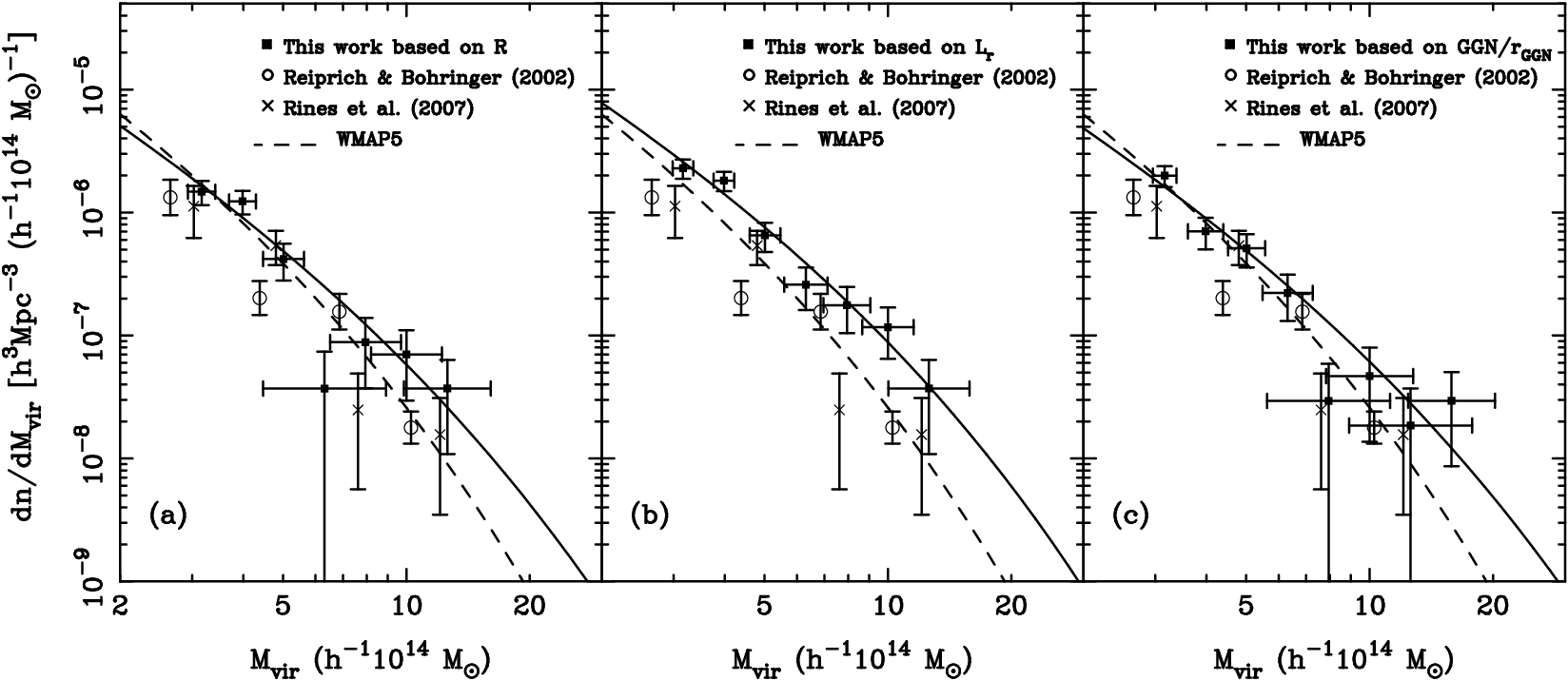}
\caption{Mass function for a sample of 56 rich clusters ($R\ge16.7$,
$0.05<z<0.1$).  The error bars on the horizontal axis are calculated
from the uncertainties of Equation~(\ref{mr})--(\ref{mggn}), and the
error bars on the vertical axis are calculated by Poisson
statistics.  The solid line is the best fit with the 
cluster mass function of Equation~(\ref{jenk2}). The dashed line is the
cluster mass function of Equation~(\ref{jenk2}) with $\Omega_m=0.273$ 
and $\sigma_8=0.813$ from the WMAP5 data \citep{kdn+09}. Data for the 
mass functions from \citet{rb02} and \citet{rdn07} are plotted for 
comparison.}
\label{mf}
\end{figure*}

\begin{figure*}
\centering
\includegraphics[width = 17.cm]{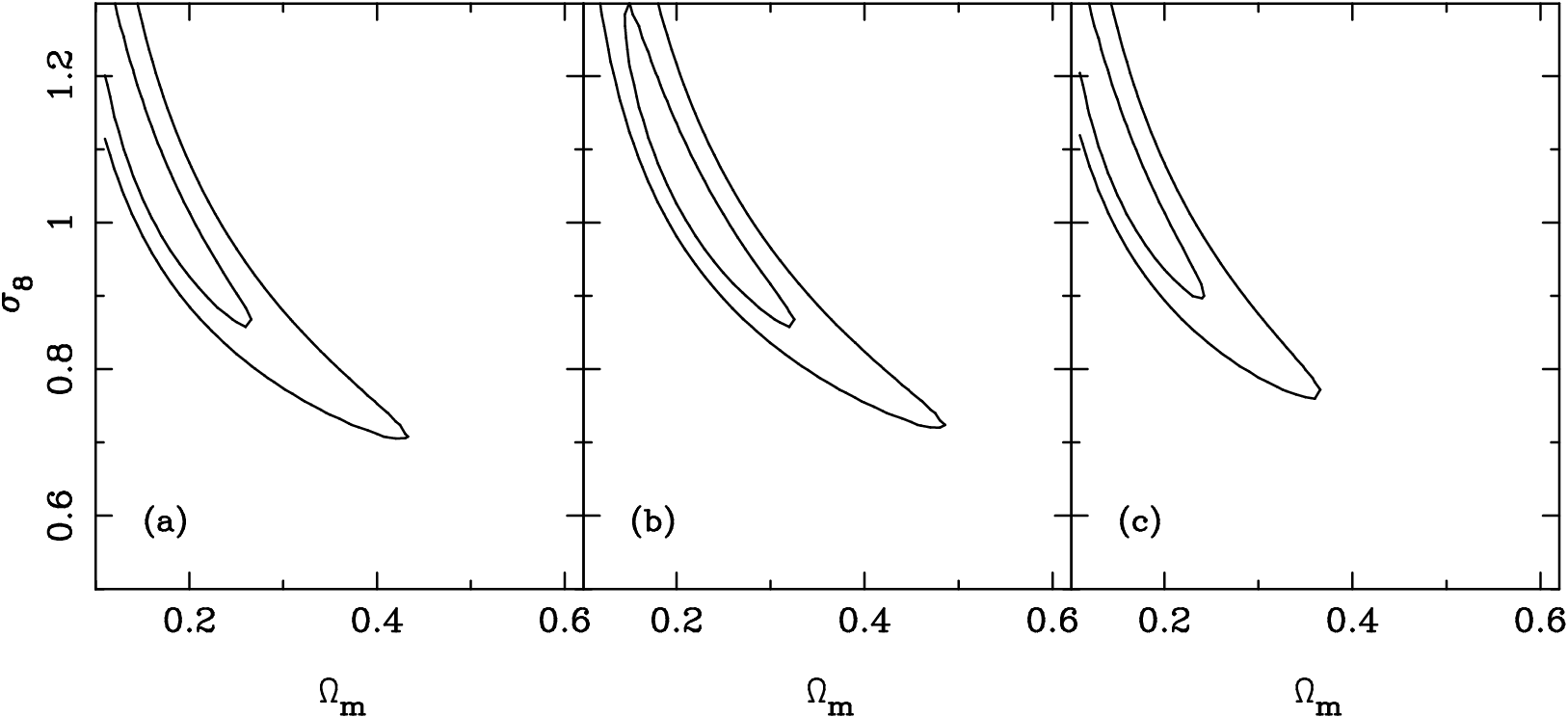}
\caption{The probability contour in the $\sigma_8$--$\Omega_m$ plane
  for three corresponding mass tracers in Figure~\ref{mf}, 68\%
  confidence level for the inner curve and 99\% for the outer curve.}
\label{cont}
\end{figure*}

First, we use a complete volume-limited sample of rich clusters ($R\ge
16.7$, 90\% complete) in the local universe ($0.05<z<0.1$) to
determine the cluster mass function. Since the photometric redshift
was used to identify the cluster member galaxies, the absolute
magnitudes of member galaxies could have large uncertainties when the
estimated cluster redshift slightly deviates from its true
redshift. To reduce the uncertainty at low redshift, we use the
spectroscopic redshifts of clusters if its discriminated members are
spectroscopically observed. The cluster richness, the summed $r$-band
luminosity and $GGN/r_{GGN}$ are re-calculated as \citet{whl09}. In this
sample, 56 clusters have richness $R\ge 16.7$, which are used to
determine the cluster mass function in the local universe.

We apply the scaling relations of Equation~(\ref{mr})--(\ref{mggn}) to
these 56 rich clusters in the local universe and calculate the number
of clusters as a function of mass. Figure~\ref{mf} shows the cluster
mass functions and the best fit with Equation~(\ref{jenk2}). From the
probability contours in the $\sigma_8$--$\Omega_m$ plane for three
mass tracers (Figure~\ref{cont}), we find that the $\sigma_8$ and
$\Omega_m$ are coupled in the form of
$\sigma_8(\Omega_m/0.3)^{\alpha}=\beta$. From the cluster mass
distribution using the mass--richness scaling relation, we find
\begin{equation}
\sigma_8\Big(\frac{\Omega_m}{0.3}\Big)^{0.42\pm0.03}=0.82\pm0.04.
\end{equation}
From the cluster mass distribution using the mass--luminosity scaling relation,
we find
\begin{equation}
\sigma_8\Big(\frac{\Omega_m}{0.3}\Big)^{0.46\pm0.03}=0.90\pm0.04.
\end{equation}
From the cluster mass distribution using the mass--$GGN/r_{GGN}$ scaling
relation, we find
\begin{equation}
\sigma_8\Big(\frac{\Omega_m}{0.3}\Big)^{0.40\pm0.03}=0.83\pm0.04.
\end{equation}
During the fitting, we have taken into account only statistical uncertainties.
Assuming $\Omega_m=0.3$, the value of $\sigma_8$ is $0.82\pm0.04$,
$0.90\pm0.04$ and $0.83\pm0.04$ for masses scaled from cluster
richness, summed luminosity and $GGN/r_{GGN}$, respectively. 

\begin{figure*}
\centering
\includegraphics[width = 17.cm]{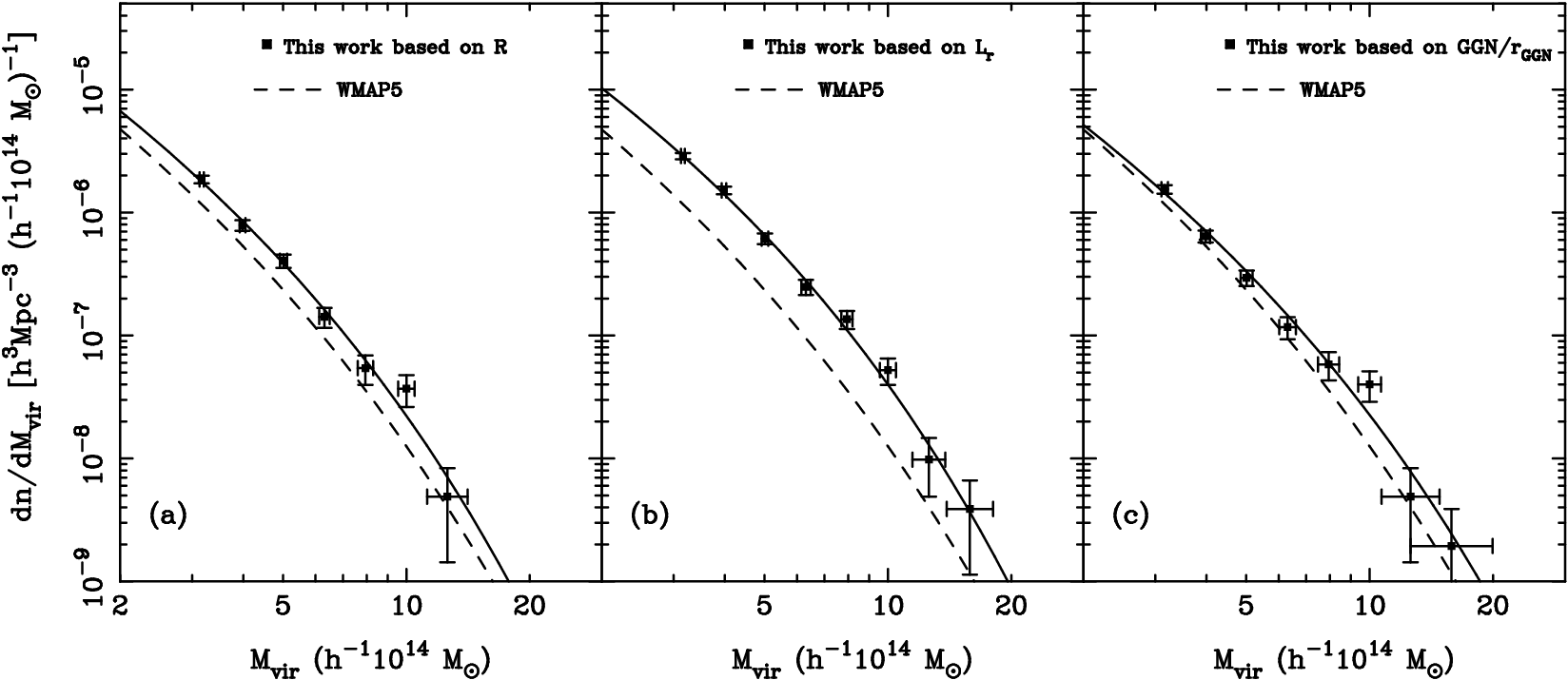}
\caption{The same as Figure~\ref{mf} but for a sample of 810 rich clusters ($R\ge16.7$, $0.2<z<0.25$).
The curve from the WMAP5 result by \citet{kdn+09} is plotted for comparison.
}
\label{mf2}
\end{figure*}

\begin{figure*}
\centering
\includegraphics[width = 17.cm]{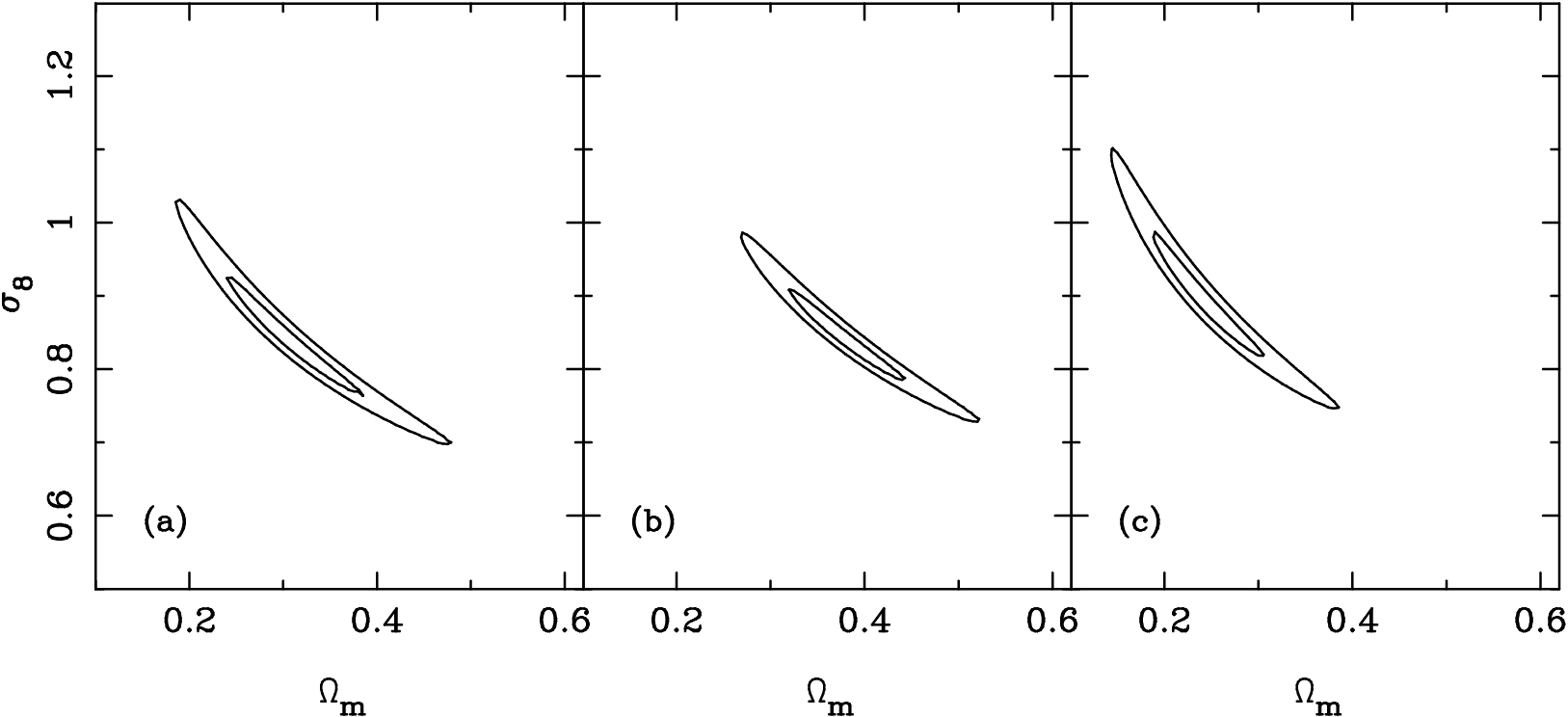}
\caption{The same as Figure~\ref{cont} but corresponding to the three mass tracers 
in Figure~\ref{mf2} for the cluster sample of $0.2<z<0.25$.}
\label{cont2}
\end{figure*}

We also apply the scaling relations of
Equation~(\ref{mr2})--(\ref{mggn2}) to the a complete volume-limited
sample of 810 rich clusters ($R\ge16.7$) of $0.2< z< 0.25$ to
calculate their masses, and get the cluster mass function.  Again,
spectroscopic redshifts of 466 clusters are used since they are
available from the SDSS, otherwise photometric redshifts are
used. Figure~\ref{mf2} shows the cluster mass functions and
Figure~\ref{cont2} shows the contours in the $\sigma_8$--$\Omega_m$
plane based on three mass tracers. Since there are much more clusters
in this sample, the mass functions have small errors than those of
$0.05<z<0.1$.  We fit the data to Equation~(\ref{jenk2}), and find
\begin{equation}
\sigma_8\Big(\frac{\Omega_m}{0.3}\Big)^{0.42\pm0.01}=0.85\pm0.02,
\end{equation}
\begin{equation}
\sigma_8\Big(\frac{\Omega_m}{0.3}\Big)^{0.46\pm0.01}=0.94\pm0.02,
\end{equation}
\begin{equation}
\sigma_8\Big(\frac{\Omega_m}{0.3}\Big)^{0.39\pm0.01}=0.82\pm0.02,
\end{equation}
for the cases using the mass tracer of richness, summed luminosity and
the $GGN/r_{GGN}$, respectively. Assuming
$\Omega_m=0.3$, the value of $\sigma_8$ is $0.85\pm0.02$,
$0.94\pm0.02$ and $0.82\pm0.02$, respectively. They are consistent
with those from the cluster sample of $0.05<z<0.1$ for each mass tracer. 

\section{Discussions and conclusions}
\label{discu}

\begin{figure}
\centering
\includegraphics[width = 7.cm]{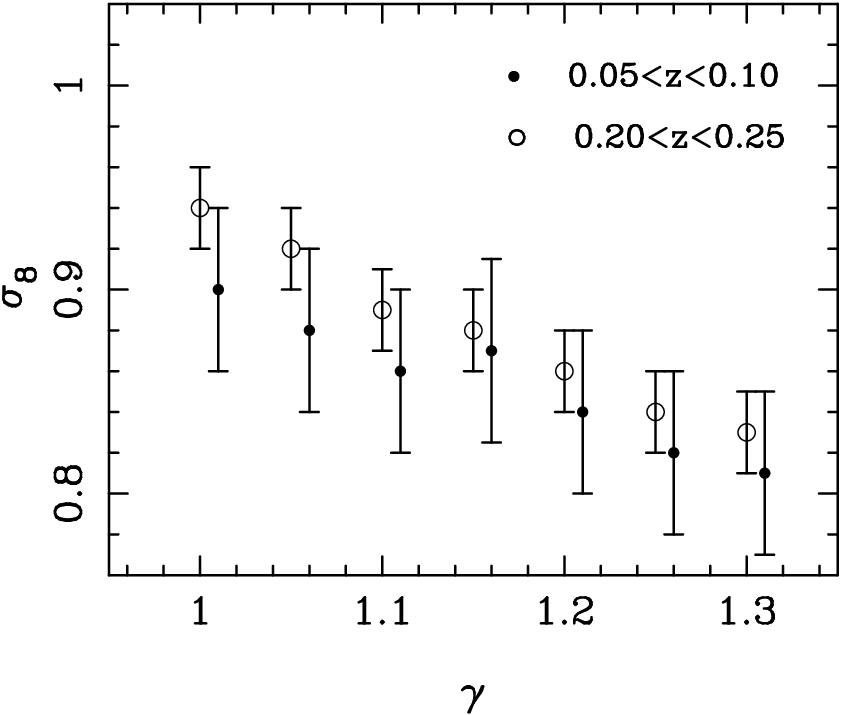}
\caption{The value of $\sigma_8$ (with $\Omega_m=0.3$ fixed) from cluster masses 
based on the $M_{\rm vir}$--$L_r$ relation varies with a possible systematic 
bias on mass conversion of $\gamma=M_{\rm vir}/M_{\rm vir, true}$.}
\label{gamma}
\end{figure}

\begin{figure}
\centering
\includegraphics[width = 7.cm]{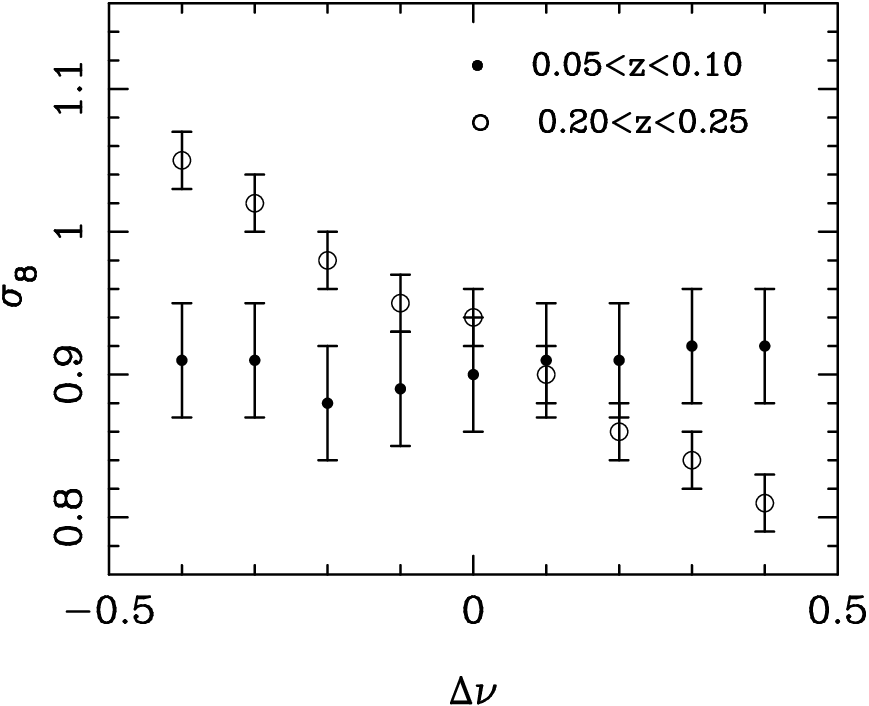}
\caption{The value of $\sigma_8$ (with $\Omega_m=0.3$ fixed) from cluster masses 
based on the $M_{\rm vir}$--$L_r$ relation varies with a possible systematic 
bias on the slope of the scaling relation by $\Delta \nu$.}
\label{slope}
\end{figure}

\begin{table*}
\begin{minipage}{160mm}
\caption[]{Comparison of results on $\sigma_8$--$\Omega_m$ derived from cluster mass function (upper part)
and cosmic microwave background (CMB) measurement (middle part).
See Table 5 of \citet{hss+07} for the results derived from weak lensing statistics.}
\begin{tabular}{lllll}
\hline
\mc{1}{c}{Reference}&\mc{1}{c}{Sample}&\mc{1}{c}{No. of clusters}&\mc{1}{c}{$\sigma_8$--$\Omega_m$ relation}& \mc{1}{c}{$\sigma_8$ ($\Omega_m=0.30$)}\\
\mc{1}{c}{}    &\mc{1}{c}{or method}&\mc{1}{c}{or observation} & \mc{1}{c}{} & \mc{1}{c}{} \\
\hline

\citet{vl96}       & X-ray   &25  & $\sigma_8=0.60\Omega_m^{-0.59+0.16\Omega_m-0.06\Omega_m^2}$          &1.16 \\
\citet{ecf96}      & X-ray   &25  & $\sigma_8$=(0.52$\pm$0.04)$\Omega_m^{-0.52+0.13\Omega_m}$          &0.93$\pm$0.07 \\ 
\citet{mar98}      & X-ray   &30  & $\sigma_8=0.78\pm0.04$ with $\Omega_m=0.30$ fixed                &0.78$\pm0.04$ \\
\citet{pen98}      & X-ray   &25  & $\sigma_8$=0.53$\Omega_m^{-0.53}$                                &1.00 \\
\citet{brt+99}     & X-ray   &70  & $\sigma_8$=0.58$\pm$0.06$\Omega_m^{-0.47\pm0.16\Omega_m}$          &0.96 \\ 
\citet{vl99}       & X-ray   &10  & $\sigma_8$=0.56$\Omega_m^{-0.47}$                                &0.99 \\
\citet{bsb+00}     & X-ray   &25  & $\sigma_8=0.96$ with $\Omega_m=0.30$ fixed                       &0.96 \\
\citet{oa01}       & X-ray   &69  & $\sigma_8$=0.59$\Omega_m^{-0.57+1.45\Omega_m-3.48\Omega_m^2+3.77\Omega_m^3-1.49\Omega_m^4}$&0.91\\ 
\citet{wu01}       & X-ray   &25  & $\sigma_8$=0.477$\Omega_m^{-0.3-0.17\Omega_m^{0.34}-0.13\Omega_{\Lambda}}$&0.87 \\
\citet{brt+01}     & X-ray   &103 & $\sigma_8$=0.66$^{+0.06}_{-0.05}$, $\Omega_m=0.35^{+0.13}_{-0.10}$ &-- \\
\citet{psw01}      & X-ray   &30  & $\sigma_8$=$(0.495^{+0.034}_{-0.037})\Omega_m^{-0.60}$             &$1.02^{+0.07}_{-0.08}$ \\
\citet{vnl02}      & X-ray   &452 & $\sigma_8$=0.38$\Omega_m^{-0.48+0.27\Omega_m}$                     &0.61 \\
\citet{rb02}       & X-ray   &106 & $\sigma_8$=0.43$\Omega_m^{-0.38}$                                &0.68\\
\citet{sel02}      & X-ray   &30  & $\sigma_8(\Omega_m/0.3)^{0.44}=0.77\pm0.07$                      &0.77$\pm$0.07 \\
\citet{vkl+03}     & X-ray   &40  & $\sigma_8=0.78^{+0.30}_{-0.06}$ with $\Omega_m=0.35$ fixed        &--\\
\citet{sbc+03}     & X-ray   &452 & $\sigma_8$=0.711$^{+0.039}_{-0.031}$, $\Omega_m=0.341^{+0.031}_{-0.029}$&--\\
\citet{pbs+03}     & X-ray   &63  & $\sigma_8$=0.77$^{+0.05}_{-0.04}$ with $\Omega_m=0.30$ fixed      &0.77$^{+0.05}_{-0.04}$\\
\citet{asf+03}     & X-ray   &111 & $\sigma_8$=(0.508$\pm$0.019)$\Omega_m^{-0.253\pm0.024}$           &0.69$\pm$0.05\\
\citet{hen04}      & X-ray   &51  & $\sigma_8=0.66\pm0.16$ with $\Omega_m=0.30$ fixed               &0.66$\pm$0.16 \\
\citet{dah06}      & X-ray   &35  & $\sigma_8(\Omega_m/0.3)^{0.37}=0.67^{+0.04}_{-0.05}$              &$0.67^{+0.04}_{-0.05}$ \\
\citet{rdn07}      & X-ray   &66  & $\sigma_8=0.92^{+0.24}_{-0.19}$, $\Omega_m=0.24^{+0.14}_{-0.09}$   &$0.84\pm0.03$\\
\citet{heh+09}     & X-ray   &48  & $\sigma_8(\Omega_m/0.32)^{0.30}=0.86\pm0.04$ for $\Omega_m\le0.32$& 0.88$\pm$0.04\\
                   &         &    & $\sigma_8(\Omega_m/0.32)^{0.41}=0.86\pm0.04$ for $\Omega_m\ge0.32$& \\
\citet{vkb+09}     & X-ray   &49  & $\sigma_8(\Omega_m/0.25)^{0.47}=0.813\pm0.027$                   &0.75$\pm$0.02\\
\citet{wef93}      & Optical &    & $\sigma_8=0.57\Omega_m^{-0.56}$                                  &1.12 \\
\citet{bf98}       & Optical & 3  & $\sigma_8\Omega_m^{0.29}=0.8\pm0.1$                              &1.13$\pm$0.14 \\
\citet{gbg+98}     & Optical &152 & $\sigma_8$=(0.60$\pm$0.04)$\Omega_m^{-0.46+0.09\Omega_m}$          &1.01$\pm$0.07 \\
\citet{bdb+03}     & Optical &300 & $\sigma_8\Omega_m^{0.60}=0.33\pm0.03$                            &0.68$\pm$0.06\\
\citet{ebc+06}     & Optical &    & $\sigma_8$=0.25$\Omega_m^{-0.92-4.5(\Omega_m-0.22)^2}$             &0.78 \\
\citet{rwr+10}     & Optical &13832 & $\sigma_8(\Omega_m/0.25)^{0.41}=0.83\pm0.03$                     &0.76$\pm$0.03\\
\hline
\citet{kdn+09}     & CMB    & WMAP5   & $\sigma_8=0.81\pm0.03$, $\Omega_m=0.27\pm0.01$    &  \\
\citet{ll09}       & CMB    & WMAP5   & $\sigma_8=0.92\pm0.04$, $\Omega_m=0.32\pm0.03$    &  \\
\citet{ldh+10}     & CMB    & WMAP7   & $\sigma_8=0.80\pm0.03$, $\Omega_m=0.26\pm0.01$    &  \\
\citet{rmc+04}     & CMB    & CBI     & $\sigma_8=0.96^{+0.06}_{-0.07}$  ~~(68\% confidence level)           & \\
\citet{dhc+06}     & CMB    & BIMA    & $\sigma_8=1.03^{+0.20}_{-0.29}$  ~~(68\% confidence level)           & \\
\citet{rab+09}     & CMB    & ACBAR   & $\sigma_8=0.93^{+0.04}_{-0.05}$            & \\
\citet{smw+09}     & CMB    & CBI     & $\sigma_8=0.92^{+0.05}_{-0.05}$            & \\
\citet{sgr+09}     & CMB    & Bolocam & $\sigma_8<1.57$ ~~~~~~~~~~~(90\% confidence level)           & \\
\citet{vac+09}     & CMB    & BOOMERANG & $\sigma_8<0.92$ ~~~~~~~~~~~(95\% confidence level)           & \\
\hline
This work ($0.05<z<0.1$)&$R$          &56 & $\sigma_8(\Omega_m/0.3)^{0.42}=0.82\pm0.04$                  &0.82$\pm$0.04\\
                        &$L_r$        &   & $\sigma_8(\Omega_m/0.3)^{0.46}=0.90\pm0.04$                  &0.90$\pm$0.04\\
                        &$GGN/r_{GGN}$&   & $\sigma_8(\Omega_m/0.3)^{0.40}=0.83\pm0.04$                  &0.83$\pm$0.04\\
This work ($0.2<z<0.25$)&$R$          &810& $\sigma_8(\Omega_m/0.3)^{0.42}=0.85\pm0.02$                  &0.85$\pm$0.02\\
                        &$L_r$        &   & $\sigma_8(\Omega_m/0.3)^{0.46}=0.94\pm0.02$                  &0.94$\pm$0.02\\
                        &$GGN/r_{GGN}$&   & $\sigma_8(\Omega_m/0.3)^{0.39}=0.82\pm0.02$                  &0.82$\pm$0.02\\
\hline
\end{tabular}
\label{comp}
\end{minipage}
\end{table*}

Cluster mass function can be accurately determined from a complete
volume-limited sample. The scaling relations of cluster mass have been
determined for three optical observations, cluster richness, summed
luminosity and $GGN/r_{GGN}$. The scaling relations are then used to
estimate cluster mass for two samples of rich clusters. We get cluster
mass functions and fit them with a theoretical expression.
Cosmological parameters are constrained in the form of
$\sigma_8(\Omega_m/0.3)^{\alpha}=\beta$, with $\alpha=$0.40--0.50 and
$\beta=$0.8--0.9. For $\Omega_m=0.3$, we get $\sigma_8=$0.8--0.9 using
different mass tracers or using the rich cluster samples at different
redshift ranges.

The $\sigma_8$ values from the mass tracers of richness $R$ and
$GGN/r_{GGN}$ obtained using both cluster samples are consistent,
while $\sigma_8$ values derived from $L_r$ are higher. This
discrepancy may come from some potential systematic bias on the mass
scaling relations. If the $M_{\rm vir}$--$L_r$ relations for both
samples are really unbiased, then the cluster masses tracer by
richness $R$ and $GGN/r_{GGN}$ are systematically
underestimated. However, it is hard to assess which one is a better
mass tracer. Given the scarce of mass estimates from different methods
for the same clusters in Table~\ref{mass} for the scaling relations,
it is also hard to estimate the systematic bias on these mass
estimates due to different methods (X-ray or weak lensing).  In our
work, one potential systematic bias may come from the conversion of
cluster mass from measured radii to the virial radius. Here, we use
$\gamma=M_{\rm vir}/M_{\rm vir, true}$ to stand for the systematic
bias of masses in Table~\ref{mass}, where $M_{\rm vir, true}$ stands
for the true virial mass of a cluster. Assuming a $\gamma$, we get
$M_{\rm vir, true}$ and then fit the mass function of clusters to
obtain $\sigma_8$. Figure~\ref{gamma} shows the variation of
$\sigma_8$ (with $\Omega_m=0.3$ fixed) as a function $\gamma$ based on
the $M_{\rm vir}$--$L_r$ relation. We are only concerned about the
cases $\gamma\ge 1$.  For example $\gamma=1.3$, i.e., masses
systematically overestimated by 30\%, the values of $\sigma_8$ are
lower by about 10\%. In fact, the deviation of $\gamma$ from 1.0 is
related to the uncertainty of intercept in the logarithm scaling
relations in Equation~(\ref{mr})--(\ref{mggn2}). The other possible
systematic bias on $\sigma_8$ may come from the slope uncertainties of
the scaling relations. Here, we illustrate the dependence of
$\sigma_8$ on the slope uncertainty, $\Delta \nu$. We only apply to
the $M_{\rm vir}$--$L_r$ relation, for example. Given a $\Delta \nu$,
i.e., $M_{\rm vir}=A\,L_r^{1+\nu+\Delta \nu}$, here $\nu=1.49$
according to Equation~(\ref{ml}) and (\ref{ml2}), we fit the power law
with the data in Figure~\ref{richmass} and \ref{richmass2} to get $A$,
and then get the cluster mass function and fit for
$\sigma_8$. Figure~\ref{slope} shows the $\sigma_8$ value varies with
$\Delta \nu$. We find that the $\sigma_8$ from the cluster sample of
$0.05<z<0.1$ does not change significantly with $\Delta \nu$, while
the $\sigma_8$ decreases from 1.05 to 0.81 for the cluster sample of
$0.2<z<0.25$ when the slope varies by $\Delta \nu$ from -0.4 to 0.4.

We can compare our results of $\sigma_8$ with previous determinations
from cluster mass function, as listed in Table~\ref{comp}. Most of
previous results are based on X-ray flux-limited cluster samples. Our
results are systematically larger than those from the mass function of
X-ray clusters.

\citet{rwr+10} used the largest number of clusters from SDSS maxBCG
catalog to determine the amplitude of cluster mass function. They did
not estimate the mass for each cluster, but gave a statistical mass
for clusters within a richness bins by weak lensing. They got
$\sigma_8=0.76\pm0.03$ assuming $\Omega_m=0.30$. The maxBCG clusters
were selected based on the red brightest cluster galaxies
(BCGs). However, the maxBCG method may miss about 25\% clusters in
which the BCGs have emission line and blue colors \citep{kma+07}. We
notice that about 15\% rich clusters ($R\ge16.7$) are missing by the
maxBCG method compared to our sample in the redshift range of $0.2<
z<0.25$. However, the systematic incompleteness only induces an
underestimate of 3\% for $\sigma_8$.  Therefore, the discrepancy
probably comes from the uncertainty of mass scaling relations. 

If we take $\Omega_m=0.26$ derived from WMAP7, then our values of
$\sigma_8$ should become larger by a factor of
$(0.26/0.3)^{\sim0.42}=1.06$, roughly equal to adding 0.05 to the
our $\sigma_8$ value in Table~\ref{comp}.  Therefore, the $\sigma_8$ values we
derived from galaxy clusters are slightly larger than the
those from the WMAP data \citep{kdn+09,ldh+10}. While
some reanalysis of the WMAP5 data independently \citep{lls+09} gives
$\sigma_8=0.921\pm0.036$ for $\Omega_m=0.32\pm0.03$ \citep[see][]{ll09}.
Some studies of cosmic microwave background at small scales also give 
higher values of $\sigma_8$ than that from WMAP \citep{rmc+04, 
dhc+06,rab+09,smw+09}. 

Our result of $\sigma_8$ are consistent with many recent studies using
other methods. For example, the $\sigma_8$ by weak lensing method has
a mean value of 0.85$\pm$0.03 \citep[see previous results in Table 5
  of ][]{hss+07}, which is higher than previous results from X-ray
clusters. \citet{tbs+04} studied the power spectrum of galaxies from
the SDSS to constrain cosmological parameters. They obtained
$\sigma_8=0.89\pm0.02$ and $\Omega_m=0.30\pm0.03$. \citet{lee09}
studied the normalization of the power spectrum via the ellipticity
function of giant galaxy voids from SDSS DR5 and obtained
$\sigma_8=0.90\pm0.04$. \citet{jnt+05} used the Ly$\alpha$ data and
found $\sigma_8=0.9$ and $\Omega_m=0.27$. \citet{fjf+03} used the
galaxy peculiar velocities to probe the growth rate of the structure
and found that $\sigma_8=1.13^{+0.22}_{-0.23}$ and
$\Omega_m=0.30^{+0.17}_{-0.07}$. 

In this work, we get six values of $\sigma_8$ by cluster mass
function. Basically, the results are consistent. However, the precise
value of $\sigma_8$ is still to be determined since our constraint is
not only coupled with $\Omega_m$, but also has large uncertainties on
the scaling relations.

\section*{Acknowledgments}

We thank the referee Dr. Vincent Eke for helpful comments.
The authors are supported by the National Natural Science Foundation
(NNSF) of China (10773016, 10821061, and 1083303) and the National Key
Basic Research Science Foundation of China (2007CB815403) and the
Liaoning Educational Fundation of China (2009A646,XN200902,054-55440105020).
Funding for the SDSS and SDSS-II has been provided by the Alfred
P. Sloan Foundation, the Participating Institutions, the National
Science Foundation, the U.S. Department of Energy, the National
Aeronautics and Space Administration, the Japanese Monbukagakusho, the
Max Planck Society, and the Higher Education Funding Council for
England. The SDSS Web site is http://www.sdss.org/.
The SDSS is managed by the Astrophysical Research Consortium for the
Participating Institutions. The Participating Institutions are the
American Museum of Natural History, Astrophysical Institute Potsdam,
University of Basel, Cambridge University, Case Western Reserve
University, University of Chicago, Drexel University, Fermilab, the
Institute for Advanced Study, the Japan Participation Group, Johns
Hopkins University, the Joint Institute for Nuclear Astrophysics, the
Kavli Institute for Particle Astrophysics and Cosmology, the Korean
Scientist Group, the Chinese Academy of Sciences (LAMOST), Los Alamos
National Laboratory, the Max Planck Institute for Astronomy (MPIA),
the Max Planck Institute for Astrophysics (MPA), New Mexico State
University, Ohio State University, University of Pittsburgh,
University of Portsmouth, Princeton University, the United States
Naval Observatory, and the University of Washington.

\label{lastpage}
\end{document}